\documentclass[conference]{IEEEtran}
\usepackage{amssymb}
\usepackage{graphicx}
\usepackage{caption}
\usepackage{subcaption}
\usepackage{tikz}
\usetikzlibrary{shapes.geometric, arrows}
\usepackage{pgfplots}
\pgfplotsset{compat=newest}
\usetikzlibrary{plotmarks}
\usepackage{grffile}
\usepackage{amsmath}
\usepackage{array} 
\usepackage{booktabs}
\usepackage{todonotes}
\usepackage{subcaption}
\DeclareMathOperator*{\argmin}{argmin}

\begin{document}
%
\title{Localization in Long-range Ultra Narrow Band IoT Networks using RSSI}

\author{\IEEEauthorblockN{
		Hazem Sallouha, Alessandro Chiumento, Sofie Pollin
		}
\IEEEauthorblockA{Department of Electrical Engineering - TELEMIC, KU Leuven\\
Kasteelpark Arenberg 10, Heverlee B-3001, Belgium\\
Email: \{hazem.sallouha, alessandro.chiumento, sofie.pollin\}@esat.kuleuven.be
}
}


%


\maketitle

\begin{abstract}
Internet of things wireless networking with long-range, low power and low throughput is raising as a new paradigm enabling to connect trillions of devices efficiently. In such networks with low power and bandwidth devices, localization becomes more challenging. In this work we take a closer look at the underlying aspects of received signal strength indicator (RSSI) based localization in UNB long-range IoT networks such as Sigfox. Firstly, the RSSI has been used for fingerprinting localization where RSSI measurements of GPS anchor nodes have been used as landmarks to classify other nodes into one of the GPS nodes classes. Through measurements we show that a location classification accuracy of 100\% is achieved when the classes of nodes are isolated. When classes are approaching each other, our measurements show that we can still achieve an accuracy of 85\%. Furthermore, when the density of the GPS nodes is increasing, we can rely on peer-to-peer triangulation and thus improve the possibility of localizing nodes with an error less than 20m from 20\% to more than 60\% of the nodes in our measurement scenario. 90\% of the nodes is localized with an error of less than 50m in our experiment with non-optimized anchor node locations.
\end{abstract}

%
\IEEEpeerreviewmaketitle

\section{Introduction}
Wireless technologies penetrate all layers of our daily lives. Concepts such as "Internet of Things (IoT)" and "Location-Based Services" are rising as new communication paradigms. The IoT is meant to enable objects of our daily life to become an integral part of the internet by equipping them with computation and communication services \cite{iot}. In such networks location information can be exploited in different layers, from communication aided purposes to the application level where location information is needed to meaningfully interpret any physical measurements collected by sensor nodes \cite{iot}, \cite{loco5g}.

In order to enable connectivity for hundreds of nodes, the currently deployed IoT networks are using long-range, low power and low throughput communications such as Sigfox and LoRa\cite{sigfox}, \cite{lora}. These characteristics make the localization problematic. On one hand, it is too expensive to integrate a GPS receiver in each node and, on the other hand, ranging-based localization techniques lack of accuracy because of low bandwidth and long distances \cite{uwb}. While Sigfox enables reception by multiple base stations, these are typically far away, in a region where the RSSI (Received Signal Strength Indicator) resolution or sensitivity of the pathloss might not be sufficient. A localization method for IoT is introduced in \cite{zhi} which can satisfy diverse requirements for indoor and outdoor scenarios. However, long-range IoT networks have not been considered.

An alternative promising method is fingerprinting-based localization \cite{fnag}-\cite{compar}. Fingerprinting localization usually works in two phases: an offline (training) phase and an online (localization) phase. During the training phase, RSSI measurements (i.e., signatures or fingerprints) are collected at known locations and stored in a database. During the online phase, a node can be localized by comparing its real-time RSSI measurements with the entries in the database \cite{simon}. The comparison process is usually performed using machine learning algorithms. A comparison between such algorithms is presented in \cite{compar}. It has been shown that support vector machines (SVM) and decision tree J48 (DTree) are among the most accurate algorithms for this range of problems \cite{Farjow}, \cite{compar}. SVM have shown good classification properties also in cases where the training dataset and the number of features are relatively small \cite{Farjow}, \cite{Tran}. This type of classification problems is encountered in low throughput networks considered in this work. An other way to exploit the RSSI measurement is by using it for distance estimation \cite{simon},\cite{kumar},\cite{impRss}. Yet, distance estimation using RSSI yields low accuracy in long distances \cite{simon}, \cite{impRss}. However, none of the mentioned works has considered the localization problem in long-range communications and this was an open problem, to the authors' best knowledge.

The contributions of this paper are twofold; firstly a measurement based localization approach that leverages the existence of some GPS nodes (i.e., anchor nodes) by using their fingerprinting in a real Sigfox deployment is proposed and discussed. This technique suits scenarios where classes (anchored by GPS nodes) are separated, such as for instance airports or other large sites. For many practical applications, it is sufficient to know in which airport your suitcase is, as an example, and it can be assumed that classes are not overlapping or not even nearby. Secondly, enhancement of the localization accuracy by relying on peer-to-peer short range communication is introduced as Sigfox nodes, which rely on TD modems, have by default a short-range communication technology on board. This enables localization within a class, when there are multiple GPS-enables peer-to-peer nodes in the class area.

The rest of the paper is organized as follows. In Section II, we provide the system model including the standards used in the experiments. Then, a clarification of the localization problem in Ultra Narrow Band (UNB) IoT networks is presented in section III. In section IV we introduce our localization approach, that includes distance estimation for short distances and fingerprinting for long distances. We present our experiments results for a real Sigofx and TD-LAN deployment in Section V. Finally, we conclude our work in Section VI.
\section{System Model}
In this section we first introduce Sigfox as an UNB technology for low power wide area networks (LPWAN); secondly we introduce a TD network protocol to be used for local area networking (TD LAN). Nodes that have TD NEXT modems, which support communication over both Sigfox and TD LAN, are considered in this work. We consider a network of three elements:
\subsubsection{\textbf{Base stations (BS)}} Long-range IoT networks can cover an entire city with very few base stations \cite{sigfox}. These base stations can enable connectivity for millions of nodes where the outdoor coverage of a single base station is up to 40km. All base stations are connected to a centralized back-end where the data and measurements can be processed.
\subsubsection{\textbf{Nodes}} 
Nodes have TD modems which support communication over both Sigfox and TD-LAN standards in 868MHz ISM-Band with a transmit power up to 25mw. 
\subsubsection{\textbf{GPS enabled nodes}} 
Some nodes are equipped with a GPS-receiver and therefore, they can send their coordinates with messages. Moreover, these nodes can receive messages from other nodes. This feature will be used in order to increase the localization accuracy based on the fact that in peer-to-peer cases the communication range is much shorter.
\subsection{Sigfox}
Sigfox \cite{sigfox}, is a IoT network operator using UNB (100Hz) channels for transmission. Binary phase shift keying (BPSK) is used as a bandwidth efficient modulation technique with a bit rate fixed to 100 bps (payload up to 12 bytes). The bandwidth of $40$kHz is split into 400 orthogonal channels and a node selects a random one for transmission \cite{brecht}. Moreover, the ISM 869 MHz band is used for outdoor communication, therefore, the MAC layer limits the number of messages to 140 messages a day (due to power-emission regulations in the ISM band). Sigfox is a cellular based network where the uplink data flow from nodes to base stations is assumed to be 97\% of the overall traffic. However, in some cases a hybrid network can be used where nodes forward data to a gateway which, in turn, sends the data to base stations \cite{TD}. One way to implement this shorter link relay is by using TD-LAN.
\subsection{TD LAN}
Alongside the ability of sending Sigfox messages, the considered nodes are also capable of short range communication. Therefore, peer-to-peer communication can be enabled to build a TD LAN network \cite{TD}. The TD LAN is an energy-efficient local area network that enables nodes to transmit data over narrow bandwidth channels (25 kHz), with a payload up to 17 bytes per second. Time devision duplex (TDD) is used in order to save power where nodes alternately transmit and receive data packets over the same radio channel. Moreover, nodes can transmit 9600 bps while using Gaussian minimum shift keying (GMSK) modulation. TD LAN uses the same ISM 869 MHz radio band (868.0 to 869.7 MHz) as Sigfox. A star topology can be formed by fixing a certain node in receive mode while its neighbors transmit upon request. 
\section{Problem definition and motivation}
Since long-range IoT networks are cellular based, one may suggest to use multilateration ranging-based techniques for localization. Since each base station covers a relatively large area (long-range base station can cover up to 40km \cite{brecht}), using multilateration will lead to a relatively large uncertainty zone. Precisely since multilateration is a ranging based technique, the accuracy depends on the estimated distances between base stations and nodes. The distances between base station and nodes can be computed using RSSI and/or time of arrival (ToA). The Cram\'er-Rao lower bound (CRLB) of an estimated distance $\hat{d}$ derived from RSSI is provided by the following inequality \cite{uwb}
\begin{eqnarray}
\sqrt{Var(\hat{d})}\geqslant \frac{\ln 10}{10}\frac{\sigma_{sh}}{n_p}d
\label{CLrss}
\end{eqnarray}
where $Var\lbrace \cdot \rbrace$ is the variance, $d$ is the actual distance between the base station and the node, $n_p$ is the path loss exponent and $\sigma_{sh}$ is the standard deviation of the zero mean Gaussian random variable representing the log-normal channel shadowing effect. From Eq. (\ref{CLrss}) we observe that increasing the distance $d$, which is the case in long-range communication, will decease the estimation accuracy by increasing the variance.

If ToA is used in ranging-based techniques, the best achievable accuracy for a distance estimate $\hat{d}$ satisfies the following inequality \cite{uwb}
\begin{eqnarray}
\sqrt{Var(\hat{d})}\geqslant \frac{c}{2\sqrt{2} \pi \sqrt{SNR} \beta}
\label{TOA_LB}
\end{eqnarray}
where $c$ is the speed of light, $SNR$ is the signal-to-noise ratio, and $\beta$ is the effective signal bandwidth. Unlike RSSI ranging technique, the accuracy of ToA depends on the signal bandwidth. In contrast, long-range communication is mainly based on ultra-narrow-band (UNB) transmission \cite{sigfox}. Therefore, this UNB (100Hz) property leads to extremely low time resolution (1-2 seconds per message) which in turn degrades the localization accuracy.
\section{Localization in UNB IoT networks}
In this section we present a long-range localization approach, which consists of two steps: region partition and localization upgrading. The basic idea of this approach is to split the wide region between transmitter and receiver into smaller classes using some GPS nodes where other nodes can be classified into one of these classes using fingerprinting classification. Afterwards, within each class, we upgrade the location accuracy using a regression process for distance estimation.
\begin{figure}[h]
	\centering
	\input{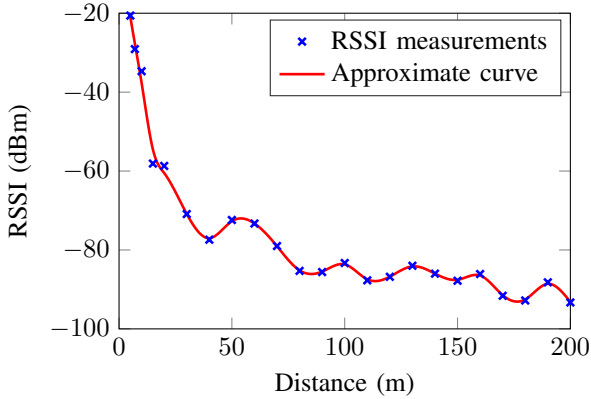}
	\vspace{-0.2cm}
	\caption{RSSI as a function of distance (m)}
	\vspace{-0.2cm}
	\label{point}
\end{figure}
The limitations of both time-based and RSSI based localization have been mentioned in the previous section. Nevertheless, RSSI has the important characteristic of presenting distinctly different behavior when the receiver is placed close to the transmitter, rather than further away as shown in Fig. \ref{point}. From this measurement-based figure it is visible that for very short distances the RSSI curve changes rapidly while as the distance keep increasing ($>$100m) the slope decreases asymptotically since the distance's influence on the RSSI starts decreasing. 
RSSI measurements can then be used in two different ways for localization purposes, distance estimation and fingerprinting. For short distances, a regression process can give a good estimation of the separation between the base station and the node as small variations in distance bring high change in RSSI values. On the other hand, as shown in Fig. \ref{point}, there are some fluctuation in the RSSI curve as distance increases. It is then possible to leverage these variations to distinguish between positions by classifying the nodes' locations based on other nodes' fluctuations in RSSI values. Fingerprinting is thus a classification framework which allows to separate the nodes in different classes by using some anchor nodes with known location (e.g., the GPS nodes).

\subsection{Fingerprinting - Classification} 
In this work, the objective is to exploit the GPS nodes as fingerprinting devices, therefore, messages sent from GPS nodes are used as landmarks (training data).
\begin{table}[h!]
	    \small
		\caption{RSSI collected by GPS nodes at a given location $l$}
		\label{rssT}
		\centering
\begin{tabular}{ccccc}
	\firsthline \hline
	\multicolumn{5}{c}{\hspace{2cm}RSSI (dBm)} \\
	\cline{2-5}
	{Time index}    & BS$^{(1)}$         & BS$^{(2)}$      &\ldots  & BS$^{(N)}$\\ 
	\hline
	t = 1           & $RSSI_1^{(1)}$    & $RSSI_1^{(2)}$ &\ldots  & $RSSI_1^{(N)}$   \\ [1ex]
	\vdots          & \vdots		   & \vdots        &$\ddots$	& \vdots 	  \\ [1ex]
	t = T           & $RSSI_T^{(1)}$    & $RSSI_T^{(2)}$ &\ldots  & $RSSI_T^{(N)}$   \\ [1ex]
	\lasthline
\end{tabular}
\end{table}
\\
The RSSI sample collected by a given GPS node at base station $n$ at time index $t$ is given by $RSSI_t^{(n)}$ where $n = 1,2,...,N$ and $t = 1,2,...,T$. Table \ref{rssT} presents RSSI measurements in a $T\times N$ matrix for a given location $l$ Assume $L$ GPS nodes at $L$ different locations, the measurements can be stored in a three dimensional matrix $T \times N \times L$. In a classification problem model this gives us a $T \times L$ set of training examples with $N$ different features. An efficient way to solve this kind of classification problems is by using SVM \cite{Farjow}.

SVM is a classification algorithm with two main components: a kernel function and a set of support vectors \cite{Tran}. The support vectors are obtained via the training phase given the training data and are chosen in order to maximize decision margins. Real-time data is classified using a simple calculation engaging the kernel function and support vectors only. The kernel function considered in this work is the Gaussian kernel due to its empirical effectiveness \cite{Farjow}, \cite{kumar}. Let $C$ denote the class of a given GPS node and $C'$ indicate the class to be determined for some test examples. The classes $C$ and $C'$ are represented by a set of RSSI measurements. Consequently, the Gaussian kernel function can be written as:
\begin{eqnarray}
K({C},{C}')&=&\exp \Big(-\frac{\lVert {C}-{C}'\lVert^2}{2\sigma^2} \Big)
\end{eqnarray}
where $\lVert.\lVert$ is the $l_2$ norm. $\sigma^2$ is the variance of the kernel function which characterizes the smoothness of the function. High $\sigma^2$ implies that the RSSI values are varying smoothly allowing more possible values for a given set of RSSI. On the other hand, low $\sigma^2$ leads to less smooth variation in the considered set of RSSI values within a given class. 
\begin{figure}[h]
	\centering
%
%
\definecolor{mycolor1}{rgb}{0.00000,0.44700,0.74100}%
\definecolor{mycolor2}{rgb}{0.85000,0.32500,0.09800}%
\definecolor{mycolor3}{rgb}{0.92900,0.69400,0.12500}%
\definecolor{mycolor4}{rgb}{0.49400,0.18400,0.55600}%
\definecolor{mycolor5}{rgb}{0.46600,0.67400,0.18800}%
\begin{tikzpicture}

\begin{axis}[%
width=6cm,
height=4.3cm,
at={(1.154in,0.752in)},
scale only axis,
xmin=-105,
xmax=-70,
xlabel={RSSI (dBm)},
ymin=0,
ymax=40,
ylabel={count},
axis background/.style={fill=white},
legend style={at={(0.03,0.97)},anchor=north west,legend cell align=left,align=left,draw=white!15!black}
]
\addplot[fill=mycolor1,fill opacity=0.6,draw=black,ybar interval,area legend] plot table[row sep=crcr] {%
x	y\\
-72.5	31\\
-71.5	29\\
-70.5	29\\
};
\addlegendentry{d = 10m};

\addplot[fill=mycolor2,fill opacity=0.6,draw=black,ybar interval,area legend] plot table[row sep=crcr] {%
x	y\\
-77.5	5\\
-76.5	36\\
-75.5	19\\
-74.5	19\\
};
\addlegendentry{d = 30m};

\addplot[fill=mycolor3,fill opacity=0.6,draw=black,ybar interval,area legend] plot table[row sep=crcr] {%
x	y\\
-87.5	14\\
-86.5	14\\
-85.5	21\\
-84.5	11\\
-83.5	11\\
};
\addlegendentry{d = 60m};

\addplot[fill=mycolor4,fill opacity=0.6,draw=black,ybar interval,area legend] plot table[row sep=crcr] {%
x	y\\
-93.5	2\\
-92.5	16\\
-91.5	14\\
-90.5	22\\
-89.5	6\\
-88.5	6\\
};
\addlegendentry{d = 80m};

\addplot[fill=mycolor5,fill opacity=0.6,draw=black,ybar interval,area legend] plot table[row sep=crcr] {%
x	y\\
-102.5	1\\
-101.5	11\\
-100.5	14\\
-99.5	24\\
-98.5	7\\
-97.5	3\\
-96.5	3\\
};
\addlegendentry{d = 130m};

\end{axis}
\end{tikzpicture}%
	\vspace{-0.2cm}
	\caption{Histogram of RSSI measurements at different distances from the receiver}
	\vspace{-0.2cm}
	\label{allvsall}
\end{figure}
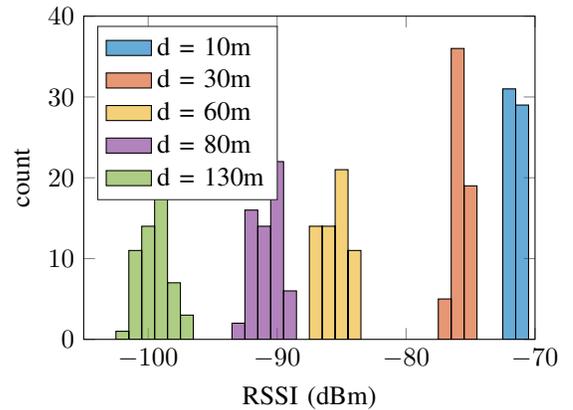
\\
Considering the RSSI histogram shown in Fig. \ref{allvsall}, a kernel function with low $\sigma^2$ ($\approx0.001$) is suitable for short distances where the RSSI is varying sharply (e.g., d = 10m and d = 30m). On the contrary, a kernel function with high $\sigma^2$ ($\approx 2$) is reasonable for longer distances where the RSSI values are varying smoothly. However, it is worth to mention that increasing $\sigma^2$ will lead to a SVM with a simple decision boundary while decreasing $\sigma^2$ gives a SVM with a complex decision boundary.

Intuitively, the RSSI data collected by nodes in an unknown class $C$ will be classified to one of the known GPS nodes classes. This means that when an error occurs, the minimum error is equal or greater than the minimum distance between all known classes with values given by
\begin{eqnarray}
\boldsymbol{e}&=&  {\lVert\hat{C}} - {{C}}\rVert^2
\label{classError}
\end{eqnarray}
where the distance $D$ between two classes (i.e., distance between two GPS nodes) is the minimum possible error
\begin{eqnarray}
D \hspace{0.2cm}=\hspace{0.2cm} \min \hspace{0.1cm}\boldsymbol{e}&=&  \min_{i,j, i\neq j}{\lVert{C_i}} - {{C_j}}\rVert^2
\label{classMinE}
\end{eqnarray}
If the average distance between any node and the GPS node in a given class is $R$, then, we need $ D\gg R$ in order to minimize the classification error.

\subsection{Distance estimation - Regression}
A further step to exploit the benefits of having GPS nodes is to use them for peer-to-peer communication with other non-GPS nodes. This can be done by building a radio map for distance estimation \cite{simon}, \cite{kumar}. In other words, instead of only classifying the nodes to one class of the GPS nodes we can further estimate the location of the node within the class by estimating the distances between the node and different GPS nodes. While this process requires at least 3 GPS nodes in each class, it can increase the localization accuracy considerably. To reduce the cost and the effort of collecting a full radio map, the regression process can be used to estimate the distance as a function of RSSI. What a regression process does is interpolate discrete data and generate a continuous output function that can be used to estimate the distance from any given RSSI value. Given the RSSI samples at $L$ different distances from every GPS node, one can model the regression problem as
\begin{eqnarray}
\argmin_{\boldsymbol{a}} \bigg\{  \Big( \sum_{l=1}^{L} {\varPsi} ({{RSSI}_l}) - {d}_l \Big)^2 \bigg\}
\label{regrProb}
\end{eqnarray}
where $d_l$ is the distance of the $l$-th location from the base station. Assume polynomial regression is used we have
\begin{eqnarray}
{\varPsi} ({{RSSI}_l})
&=& a_0 + a_1{RSSI}_l^{} + a_2{RSSI}_l^{2} + \nonumber\\
 &\cdots& + a_N{RSSI}_l^{n}
\label{regrPoly}
\end{eqnarray}
where $a_j$ (with $j = 0, 1, 2, ..., n$) are the coefficients of the polynomial and $l = 1,2,...,L$. Polynomial regression has been chosen since polynomials dominate the interpolation theory$-$Weierstrass's theorem and they are easy to evaluate \cite{impRss}. In order to overcome the effect of small scale fading we have collected multiple RSSI measurements from every location and averaged them over time. Once ${\varPsi} ({{RSSI}_l})$ has been defined, the estimated distance of a new $RSSI_m$ value at unknown distance is given by
\begin{eqnarray}
{\hat{d}} &=& \argmin_{d_l} \bigg\{  \Big( {\varPsi}^{-1} (d_l) - {RSSI}_m \Big)^2 \bigg\}
\label{regrDis}
\end{eqnarray}
However, the function ${\varPsi} ({{RSSI}_l})$ will not fit the data perfectly. Thus, an estimation error is expected. In case of multiple test data of length $M$. The RMS error for an error vector of length $M$ is defined as
\begin{eqnarray}
{e_{rms}}&=& \sqrt{\frac{1}{M}\sum_{m=1}^{M}{{\rVert \hat{d}_m} - {d_m} \rVert}^2}
\label{regrRMSE}
\end{eqnarray}
This is for a distance estimation between a given node and one GPS node. In order to estimate the position of a node using distance estimation we need at least 3 GPS nodes. Note that in case of regression the function ${\varPsi}^{-1} (d_l)$ is continuous. Therefore, the estimated distance $\hat{d}$ can take any non-negative value.
\section{Experiments and Results}
In this section experiments using nodes that support communication over both Sigfox and TD-LAN are presented. The experiments have been conducted in two different scenarios. In the first one, we only consider communication over Sigfox network. The object of this scenario is to use the GPS nodes to split the wide coverage area into smaller classes. In the second scenario we enabled the peer-to-peer communication over TD-LAN network to increase the localization accuracy. 
\subsection{Communication over Sigfox}
\begin{figure}
	\centering
	\begin{subfigure}{.24\textwidth}
		\centering
		\includegraphics[width=3.8cm,height=6cm]{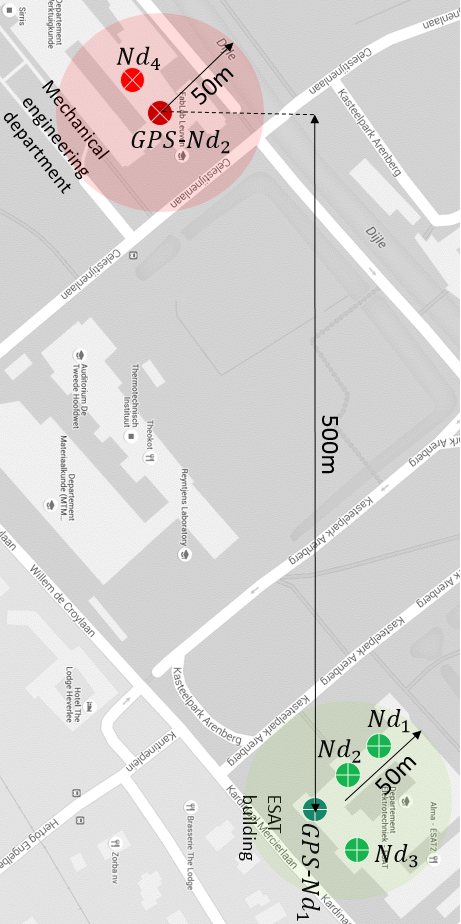}
		\caption{$D = 10R$}
		\label{map}
	\end{subfigure}%
	\begin{subfigure}{.28\textwidth}
		\centering
		\includegraphics[width=3.8cm,height=6cm]{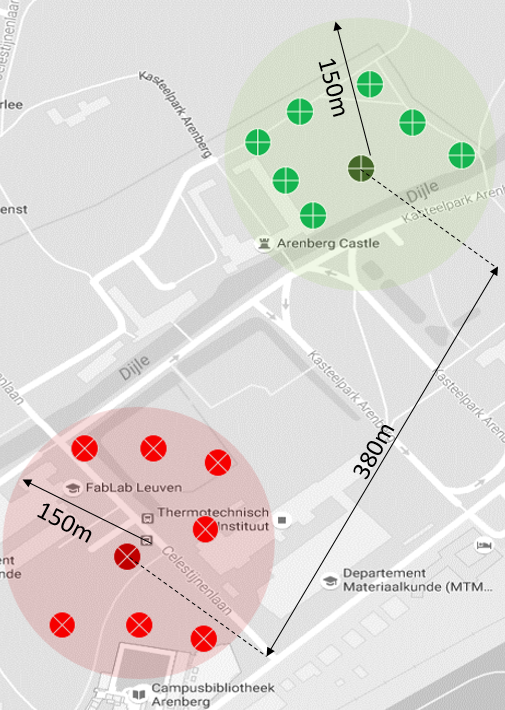}
		\caption{$D= 2.5R$}
		\label{map2}
	\end{subfigure}
	\caption{Map with nodes' positions}
	\label{fig:test}
	\vspace{-0.2cm}
\end{figure}

In this scenario we investigated the achieved localization accuracy using Sigfox network where all nodes directed their messages to the base stations. In the first test setup six nodes in different positions have been used where each node has sent 100 messages. A map of the nodes positioned on top of two different buildings in Arenberg campus (i.e., two different classes) is shown in Fig. \ref{map}.

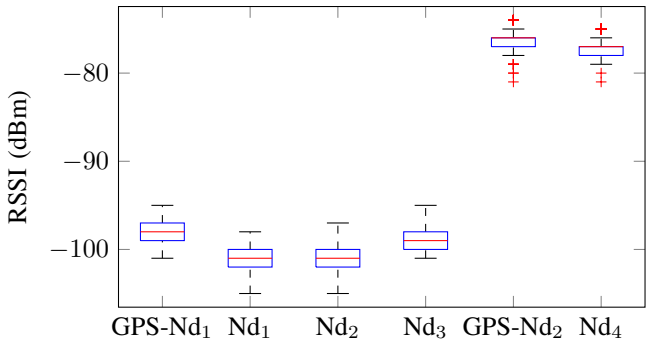
\begin{figure}[h]
	\centering
%
%
\begin{tikzpicture}

\begin{axis}[%
width=7cm,
height=4cm,
at={(0.758in,0.481in)},
scale only axis,
unbounded coords=jump,
xmin=0.5,
xmax=6.5,
xtick={1,2,3,4,5,6},
ylabel={RSSI (dBm)},
xticklabels={{GPS-Nd$_1$},{Nd$_1$},{Nd$_2$},{Nd$_3$},{GPS-Nd$_2$},{Nd$_4$}},
ymin=-106.55,
ymax=-72.45,
axis background/.style={fill=white}
]
\addplot [color=black,dashed,forget plot]
  table[row sep=crcr]{%
1	-97\\
1	-95\\
};
\addplot [color=black,dashed,forget plot]
  table[row sep=crcr]{%
2	-100\\
2	-98\\
};
\addplot [color=black,dashed,forget plot]
  table[row sep=crcr]{%
3	-100\\
3	-97\\
};
\addplot [color=black,dashed,forget plot]
  table[row sep=crcr]{%
4	-98\\
4	-95\\
};
\addplot [color=black,dashed,forget plot]
  table[row sep=crcr]{%
5	-76\\
5	-75\\
};
\addplot [color=black,dashed,forget plot]
  table[row sep=crcr]{%
6	-77\\
6	-76\\
};
\addplot [color=black,dashed,forget plot]
  table[row sep=crcr]{%
1	-101\\
1	-99\\
};
\addplot [color=black,dashed,forget plot]
  table[row sep=crcr]{%
2	-105\\
2	-102\\
};
\addplot [color=black,dashed,forget plot]
  table[row sep=crcr]{%
3	-105\\
3	-102\\
};
\addplot [color=black,dashed,forget plot]
  table[row sep=crcr]{%
4	-101\\
4	-100\\
};
\addplot [color=black,dashed,forget plot]
  table[row sep=crcr]{%
5	-78\\
5	-77\\
};
\addplot [color=black,dashed,forget plot]
  table[row sep=crcr]{%
6	-79\\
6	-78\\
};
\addplot [color=black,solid,forget plot]
  table[row sep=crcr]{%
0.875	-95\\
1.125	-95\\
};
\addplot [color=black,solid,forget plot]
  table[row sep=crcr]{%
1.875	-98\\
2.125	-98\\
};
\addplot [color=black,solid,forget plot]
  table[row sep=crcr]{%
2.875	-97\\
3.125	-97\\
};
\addplot [color=black,solid,forget plot]
  table[row sep=crcr]{%
3.875	-95\\
4.125	-95\\
};
\addplot [color=black,solid,forget plot]
  table[row sep=crcr]{%
4.875	-75\\
5.125	-75\\
};
\addplot [color=black,solid,forget plot]
  table[row sep=crcr]{%
5.875	-76\\
6.125	-76\\
};
\addplot [color=black,solid,forget plot]
  table[row sep=crcr]{%
0.875	-101\\
1.125	-101\\
};
\addplot [color=black,solid,forget plot]
  table[row sep=crcr]{%
1.875	-105\\
2.125	-105\\
};
\addplot [color=black,solid,forget plot]
  table[row sep=crcr]{%
2.875	-105\\
3.125	-105\\
};
\addplot [color=black,solid,forget plot]
  table[row sep=crcr]{%
3.875	-101\\
4.125	-101\\
};
\addplot [color=black,solid,forget plot]
  table[row sep=crcr]{%
4.875	-78\\
5.125	-78\\
};
\addplot [color=black,solid,forget plot]
  table[row sep=crcr]{%
5.875	-79\\
6.125	-79\\
};
\addplot [color=blue,solid,forget plot]
  table[row sep=crcr]{%
0.75	-99\\
0.75	-97\\
1.25	-97\\
1.25	-99\\
0.75	-99\\
};
\addplot [color=blue,solid,forget plot]
  table[row sep=crcr]{%
1.75	-102\\
1.75	-100\\
2.25	-100\\
2.25	-102\\
1.75	-102\\
};
\addplot [color=blue,solid,forget plot]
  table[row sep=crcr]{%
2.75	-102\\
2.75	-100\\
3.25	-100\\
3.25	-102\\
2.75	-102\\
};
\addplot [color=blue,solid,forget plot]
  table[row sep=crcr]{%
3.75	-100\\
3.75	-98\\
4.25	-98\\
4.25	-100\\
3.75	-100\\
};
\addplot [color=blue,solid,forget plot]
  table[row sep=crcr]{%
4.75	-77\\
4.75	-76\\
5.25	-76\\
5.25	-77\\
4.75	-77\\
};
\addplot [color=blue,solid,forget plot]
  table[row sep=crcr]{%
5.75	-78\\
5.75	-77\\
6.25	-77\\
6.25	-78\\
5.75	-78\\
};
\addplot [color=red,solid,forget plot]
  table[row sep=crcr]{%
0.75	-98\\
1.25	-98\\
};
\addplot [color=red,solid,forget plot]
  table[row sep=crcr]{%
1.75	-101\\
2.25	-101\\
};
\addplot [color=red,solid,forget plot]
  table[row sep=crcr]{%
2.75	-101\\
3.25	-101\\
};
\addplot [color=red,solid,forget plot]
  table[row sep=crcr]{%
3.75	-99\\
4.25	-99\\
};
\addplot [color=red,solid,forget plot]
  table[row sep=crcr]{%
4.75	-76\\
5.25	-76\\
};
\addplot [color=red,solid,forget plot]
  table[row sep=crcr]{%
5.75	-77\\
6.25	-77\\
};
\addplot [color=black,only marks,mark=+,mark options={solid,draw=red},forget plot]
  table[row sep=crcr]{%
nan	nan\\
};
\addplot [color=black,only marks,mark=+,mark options={solid,draw=red},forget plot]
  table[row sep=crcr]{%
nan	nan\\
};
\addplot [color=black,only marks,mark=+,mark options={solid,draw=red},forget plot]
  table[row sep=crcr]{%
nan	nan\\
};
\addplot [color=black,only marks,mark=+,mark options={solid,draw=red},forget plot]
  table[row sep=crcr]{%
nan	nan\\
};
\addplot [color=black,only marks,mark=+,mark options={solid,draw=red},forget plot]
  table[row sep=crcr]{%
5	-81\\
5	-80\\
5	-80\\
5	-79\\
5	-79\\
5	-79\\
5	-79\\
5	-79\\
5	-79\\
5	-79\\
5	-74\\
5	-74\\
5	-74\\
5	-74\\
5	-74\\
5	-74\\
5	-74\\
5	-74\\
5	-74\\
5	-74\\
5	-74\\
5	-74\\
5	-74\\
5	-74\\
5	-74\\
5	-74\\
5	-74\\
5	-74\\
5	-74\\
5	-74\\
5	-74\\
5	-74\\
5	-74\\
5	-74\\
5	-74\\
5	-74\\
};
\addplot [color=black,only marks,mark=+,mark options={solid,draw=red},forget plot]
  table[row sep=crcr]{%
6	-81\\
6	-80\\
6	-75\\
6	-75\\
6	-75\\
6	-75\\
6	-75\\
6	-75\\
6	-75\\
6	-75\\
6	-75\\
6	-75\\
6	-75\\
6	-75\\
6	-75\\
6	-75\\
6	-75\\
6	-75\\
6	-75\\
6	-75\\
6	-75\\
6	-75\\
6	-75\\
6	-75\\
6	-75\\
6	-75\\
6	-75\\
};
\end{axis}
\end{tikzpicture}%
	\vspace{-0.2cm}
	\caption{The characteristics of RSSI measurements vs node ID}
	\vspace{-0.6cm}
	\label{exp1}
\end{figure}
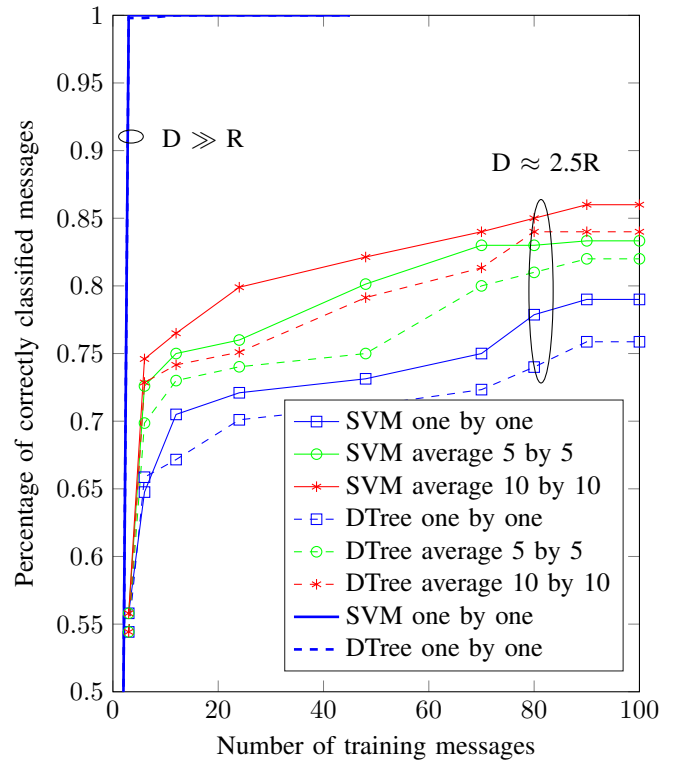
\begin{figure}[t]
	\centering
%
%
\begin{tikzpicture}

\begin{axis}[%
width=7cm,
height=9cm,
at={(0.758in,0.481in)},
scale only axis,
xmin=0,
xmax=100,
xlabel={Number of training messages},
ymin=0.5,
ymax=1.0,
ylabel={Percentage of correctly classified messages},
axis background/.style={fill=white},
legend style={at={(0.97,0.03)},anchor=south east,legend cell align=left,align=left,draw=white!15!black}
]
\addplot [color=blue,solid,mark=square,mark options={solid}]
  table[row sep=crcr]{%
3	0.558\\
6	0.6476\\
12	0.705\\
24	0.721\\
48	0.73133\\
70	0.75\\
80	0.7787\\
90	0.79\\
100	0.79\\
};
\addlegendentry{SVM one by one};

\addplot [color=green,solid,mark=o,mark options={solid}]
  table[row sep=crcr]{%
3	0.558\\
6	0.726\\
12	0.75\\
24	0.76\\
48	0.80133\\
70	0.83\\
80	0.83\\
90	0.8333\\
100	0.8333\\
};
\addlegendentry{SVM average 5 by 5};

\addplot [color=red,solid,mark=asterisk,mark options={solid}]
  table[row sep=crcr]{%
3	0.558\\
6	0.746\\
12	0.765\\
24	0.799\\
48	0.82133\\
70	0.84\\
80	0.85\\
90	0.86\\
100	0.86\\
};
\addlegendentry{SVM average 10 by 10};

\addplot [color=blue,dashed,mark=square,mark options={solid}]
  table[row sep=crcr]{%
3	0.54424\\
6	0.6586\\
12	0.671619\\
24	0.70102\\
48	0.71133\\
70	0.72333\\
80	0.74\\
90	0.7587\\
100	0.7587\\
};
\addlegendentry{DTree one by one};

\addplot [color=green,dashed,mark=o,mark options={solid}]
  table[row sep=crcr]{%
3	0.54424\\
6	0.6986\\
12	0.73\\
24	0.7402\\
48	0.75\\
70	0.8\\
80	0.81\\
90	0.82\\
100	0.82\\
};
\addlegendentry{DTree average 5 by 5};

\addplot [color=red,dashed,mark=asterisk,mark options={solid}]
  table[row sep=crcr]{%
3	0.54424\\
6	0.7286\\
12	0.741619\\
24	0.75102\\
48	0.79133\\
70	0.81333\\
80	0.84\\
90	0.84\\
100	0.84\\
};
\addlegendentry{DTree average 10 by 10};

\addplot [color=blue,solid,line width=1.0pt]
table[row sep=crcr]{%
1	0\\
3	0.99999\\
6	0.99999\\
12	0.99999\\
24	0.99999\\
30		0.99999\\
35	0.99999\\
40		0.99999\\
45	0.99999\\
};
\addlegendentry{SVM one by one};

\addplot [color=blue,dashed,line width=1.0pt]
table[row sep=crcr]{%
1	0\\
3	0.998\\
6	0.998\\
12	0.99999\\
24	0.99999\\
30		0.99999\\
35	0.99999\\
40		0.99999\\
45	0.99999\\
};
\addlegendentry{DTree one by one};

\end{axis}
\begin{axis}[%
width=8cm,
height=9cm,
at={(0.758in,0.481in)},
scale only axis,
xmin=0,
xmax=1,
ymin=0,
ymax=1,
hide axis,
axis x line*=bottom,
axis y line*=left
]
\node[below, align=center, text=black]
at (rel axis cs:0.7200,0.81) {D $\approx$ 2.5R};
\node[below, align=center, text=black]
at (rel axis cs:0.15,0.845077) {D $\gg$ R};
\draw [black,solid] (axis cs:0.711462,0.5920806) ellipse [x radius=0.02012911, y radius=0.135608];
\draw [black,solid] (axis cs:0.03,0.82077) ellipse [x radius=0.0204765, y radius=0.00959488];
\end{axis}
\end{tikzpicture}%
	\vspace{-0.6cm}
	\caption{Percentage of correctly classified messages for both SVM and decision tree algorithms}
	\vspace{-0.4cm}
	\label{ave}
\end{figure}

In this setup the average distance $R$ between any node (Nd) and the GPS node (GPS-Nd) within one class is $10$m. The RSSI measurements at the nearest base station versus device ID is presented in Fig. \ref{exp1}. Obviously, since $D \gg R$ (i.e., $D = 10R$), one can see that the RSSI values of the two groups are separated by $23$ dB. With this test setup nodes can be easily classified into two classes using simple machine learning algorithms such as DTree. As shown in Fig. \ref{ave}, only 1-2 training messages from each GPS node are enough to get a 100\% correctly classification of the 100 messages from the other 3 nodes. However, as $R$ approaches $D$, the minimum number of message will increase since more training data will be needed to learn the class of each node.

\begin{figure}[h]
	\centering
%
%
\definecolor{mycolor1}{rgb}{0.00000,0.44700,0.74100}%
\definecolor{mycolor2}{rgb}{0.85000,0.32500,0.09800}%
\begin{tikzpicture}

\begin{axis}[%
width=7cm,
height=4cm,
at={(1.154in,0.752in)},
scale only axis,
xmin=0,
xmax=70,
xlabel={$\sigma{}^\text{2}$},
ymin=0.5,
ymax=0.9,
ylabel={Classification accuracy},
axis background/.style={fill=white},
legend style={legend cell align=left,align=left,draw=white!15!black}
]
\addplot [color=mycolor1,dashdotted,line width=1.0pt]
  table[row sep=crcr]{%
0.1	0.52\\
0.125	0.535833333333333\\
0.166666666666667	0.56225\\
0.25	0.58375\\
0.333333333333333	0.6115\\
0.5	0.6405\\
0.666666666666667	0.65975\\
0.833333333333333	0.691\\
1	0.7135\\
1.25	0.72975\\
1.66666666666667	0.74375\\
2	0.75325\\
2.5	0.75965\\
3.33333333333333	0.75965\\
10	0.74515\\
15.1515151515152	0.70115\\
25	0.6529\\
33.3333333333333	0.60272\\
50	0.5312\\
71.4285714285714	0.5111\\
};
\addlegendentry{SVM one-by-one};

\addplot [color=green,dashed,line width=1.0pt]
  table[row sep=crcr]{%
0.1	0.53\\
0.125	0.533333333333333\\
0.166666666666667	0.54675\\
0.25	0.56275\\
0.333333333333333	0.59275\\
0.5	0.63275\\
0.666666666666667	0.67275\\
0.833333333333333	0.714\\
1	0.75\\
1.25	0.778\\
1.66666666666667	0.796\\
2	0.808\\
2.5	0.8104\\
3.33333333333333	0.8044\\
10	0.7804\\
15.1515151515152	0.7253\\
25	0.6678\\
33.3333333333333	0.61162\\
50	0.5327\\
71.4285714285714	0.5111\\
};
\addlegendentry{SVM average 5-by-5};

\addplot [color=mycolor2,solid,line width=1.0pt]
  table[row sep=crcr]{%
0.1	0.53\\
0.125	0.5369\\
0.166666666666667	0.55011\\
0.25	0.56611\\
0.333333333333333	0.59847\\
0.5	0.63807\\
0.666666666666667	0.68065\\
0.833333333333333	0.7319\\
1	0.7819\\
1.25	0.8142\\
1.66666666666667	0.8358\\
2	0.848\\
2.5	0.8464\\
3.33333333333333	0.8244\\
10	0.78944\\
15.1515151515152	0.73034\\
25	0.66884\\
33.3333333333333	0.60666\\
50	0.5327\\
71.4285714285714	0.5111\\
};
\addlegendentry{SVM average 10-by-10};

\end{axis}
\end{tikzpicture}%
	\vspace{-0.2cm}
	\caption{The classification accuracy as a function of $\sigma^2$}
	\vspace{-0.2cm}
	\label{sigma}
\end{figure}
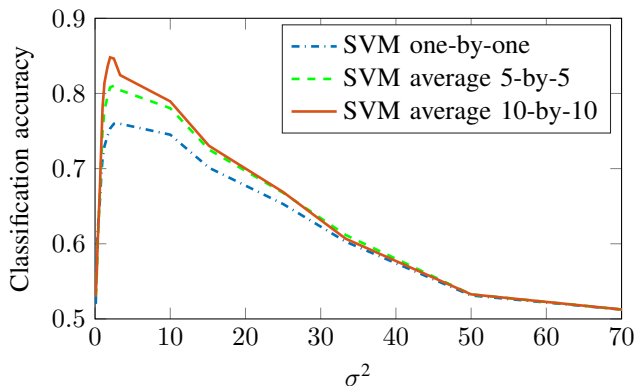

Another test setup with 16 nodes with $D \approx 2.5R$ is shown in Fig. \ref{map2}. Using this setup we can investigate the effects of node density (i.e., varying $R$ and $D$) on the classification accuracy. To this end, the nodes have been classified into the GPS nodes' classes. Each node has sent 100 messages from which the RSSI values are measured at all base stations that have received the messages. As mentioned earlier, the messages received from the GPS nodes will be used as training data in order to classify the other nodes. In this case the classification problem will be more challenging as the measurements of every group will be less correlated. Therefore, choosing $\sigma^2$ is crucial. Classification accuracy of 100 messages from each node using 100 messages from GPS nodes as training data as a function of $\sigma^2$ is illustrated in Fig. \ref{sigma}. It can be seen that $\sigma^2 \approx 4$ gives the best classification accuracy. Therefore, a kernel function with $\sigma^2 = 4$ is used in the SVM algorithm. The classification accuracy for both SVM and DTree with different number of training data is presented in Fig. \ref{ave}. As shown in the figure, the best achievable classification accuracy is 78\% when classifying messages one-by-one.

A possible improvement can be earned by decreasing the RSSI randomness that occurs due to channel fluctuation by averaging the RSSI values. By averaging the messages 10-by-10 we can get a classification accuracy of 87\% using a SVM classifier. A wrongly classified message means that the estimated location of the node will be in another class. The distance error, in this case, is the distance between the correct and the estimated class.
The fingerprinting discussed in this section is then able to give good localization accuracy when $D \gg R$. In order to improve on these results, the impact of a TD-LAN network between the nodes within each class is further presented. 
\subsection{Peer-to-Peer enabled for higher accuracy}
In case of TD-LAN enabled nodes, signal powers can be collected in the near zone where the RSSI values have better resolution. This resolution is high enough to estimate the distance of the node from its RSSI at the receiver. In order to estimate the changing of RSSI values with distance, three nodes 10m apart on a line have been used as receivers. Transmitters have been placed at different distances from 10 to 200m with increments of 10m, orthogonal to the line made by the three receivers. We note that this configuration is not optimal to cover a certain area, as this would require the receiving nodes to be placed on the corners of the considered area or line. The current results hence represent a lower bound on the accuracy.
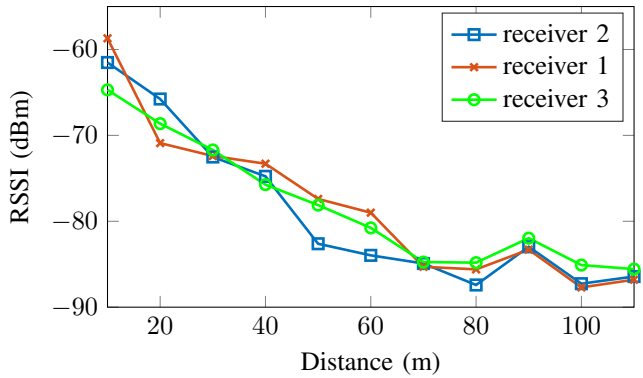
\begin{figure}[h]
	\centering
%
%
\definecolor{mycolor1}{rgb}{0.00000,0.44700,0.74100}%
\definecolor{mycolor2}{rgb}{0.85000,0.32500,0.09800}%
\begin{tikzpicture}

\begin{axis}[%
width=7cm,
height=4cm,
at={(1.154in,0.752in)},
scale only axis,
xmin=10,
xmax=110,
xlabel={Distance (m)},
ymin=-90,
ymax=-55,
ylabel={RSSI (dBm)},
axis background/.style={fill=white},
legend style={legend cell align=left,align=left,draw=white!15!black}
]
\addplot [color=mycolor1,solid,line width=1.0pt,mark=square,mark options={solid}]
  table[row sep=crcr]{%
10	-61.5166666666667\\
20	-65.7666666666667\\
30	-72.5166666666667\\
40	-74.7666666666667\\
50	-82.6166666666667\\
60	-83.9666666666667\\
70	-84.9166666666667\\
80	-87.4166666666667\\
90	-82.9666666666667\\
100	-87.2666666666667\\
110	-86.4333333333333\\
};
\addlegendentry{receiver 2};

\addplot [color=mycolor2,solid,line width=1.0pt,mark=x,mark options={solid}]
  table[row sep=crcr]{%
10	-58.7\\
20	-70.9\\
30	-72.4\\
40	-73.3\\
50	-77.4\\
60	-79\\
70	-85.3\\
80	-85.6\\
90	-83.3\\
100	-87.7\\
110	-86.8\\
};
\addlegendentry{receiver 1};

\addplot [color=green,solid,line width=1.0pt,mark=o,mark options={solid}]
  table[row sep=crcr]{%
10	-64.7\\
20	-68.6333333333333\\
30	-71.7\\
40	-75.7333333333333\\
50	-78.1166666666667\\
60	-80.7666666666667\\
70	-84.75\\
80	-84.8166666666667\\
90	-81.9666666666667\\
100	-85.1\\
110	-85.5666666666667\\
};
\addlegendentry{receiver 3};

\end{axis}
\end{tikzpicture}%
	\vspace{-0.2cm}
	\caption{Average RSSI values at three different receivers separated by 10m from each other in a TD LAN network}
	\label{3rec}
\end{figure}
\begin{figure}[h]
	\centering
	\input{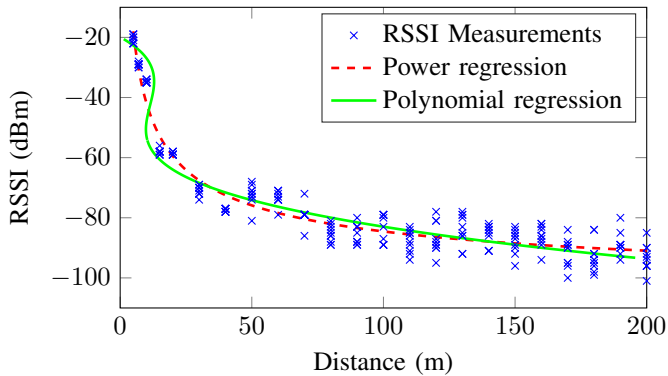}
	\vspace{-0.6cm}
	\caption{Power regression and polynomial regression for the collected data at different distances}
	\vspace{-0.4cm}
	\label{regr}
\end{figure}
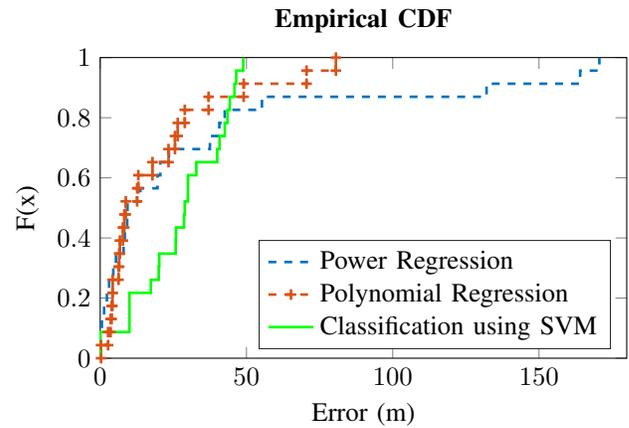
\begin{figure}[h]
	\centering
%
%
\definecolor{mycolor1}{rgb}{0.00000,0.44700,0.74100}%
\definecolor{mycolor2}{rgb}{0.85000,0.32500,0.09800}%
\begin{tikzpicture}

\begin{axis}[%
width=7cm,
height=4cm,
at={(1.387in,1.326in)},
scale only axis,
unbounded coords=jump,
xmin=0,
xmax=180,
xlabel={Error (m)},
ymin=0,
ymax=1,
ylabel={F(x)},
axis background/.style={fill=white},
title style={font=\bfseries},
title={Empirical CDF},
legend style={at={(0.97,0.03)},anchor=south east,legend cell align=left,align=left,draw=white!15!black}
]
\addplot [color=mycolor1,dashed,line width=1.0pt]
  table[row sep=crcr]{%
-inf	0\\
0.00636472565428985	0\\
0.00636472565428985	0.0434782608695652\\
0.0233883126351838	0.0434782608695652\\
0.0233883126351838	0.0869565217391304\\
0.5674595849296	0.0869565217391304\\
0.5674595849296	0.130434782608696\\
1.32432190293137	0.130434782608696\\
1.32432190293137	0.173913043478261\\
2.29893555631424	0.173913043478261\\
2.29893555631424	0.217391304347826\\
3.04949488241802	0.217391304347826\\
3.04949488241802	0.260869565217391\\
4.47043295438757	0.260869565217391\\
4.47043295438757	0.304347826086957\\
5.38115622864427	0.304347826086957\\
5.38115622864427	0.347826086956522\\
7.98296700974063	0.347826086956522\\
7.98296700974063	0.391304347826087\\
8.40167586070025	0.391304347826087\\
8.40167586070025	0.434782608695652\\
9.21266198881331	0.434782608695652\\
9.21266198881331	0.478260869565217\\
9.33288666297229	0.478260869565217\\
9.33288666297229	0.521739130434783\\
13.73047293491	0.521739130434783\\
13.73047293491	0.565217391304348\\
19.5887604445114	0.565217391304348\\
19.5887604445114	0.608695652173913\\
20.5578763125345	0.608695652173913\\
20.5578763125345	0.652173913043478\\
23.4928935828165	0.652173913043478\\
23.4928935828165	0.695652173913043\\
37.500656557522	0.695652173913043\\
37.500656557522	0.739130434782609\\
40.7209778496085	0.739130434782609\\
40.7209778496085	0.782608695652174\\
42.5776844272856	0.782608695652174\\
42.5776844272856	0.826086956521739\\
55.2888891060046	0.826086956521739\\
55.2888891060046	0.869565217391304\\
132.079076577927	0.869565217391304\\
132.079076577927	0.91304347826087\\
164.070978813975	0.91304347826087\\
164.070978813975	0.956521739130435\\
170.625004750871	0.956521739130435\\
170.625004750871	1\\
inf	1\\
};
\addlegendentry{Power Regression};

\addplot [color=mycolor2,dashed,line width=1.0pt,mark=+,mark options={solid}]
  table[row sep=crcr]{%
-inf	0\\
0.290404907000124	0\\
0.290404907000124	0.0434782608695652\\
2.69457957899996	0.0434782608695652\\
2.69457957899996	0.0869565217391304\\
3.54067703200002	0.0869565217391304\\
3.54067703200002	0.130434782608696\\
3.91329900799983	0.130434782608696\\
3.91329900799983	0.173913043478261\\
4.18574699999996	0.173913043478261\\
4.18574699999996	0.217391304347826\\
4.24918088299998	0.217391304347826\\
4.24918088299998	0.260869565217391\\
6.26658258100017	0.260869565217391\\
6.26658258100017	0.304347826086957\\
6.47980844800023	0.304347826086957\\
6.47980844800023	0.347826086956522\\
6.72793369599978	0.347826086956522\\
6.72793369599978	0.391304347826087\\
7.74448589900001	0.391304347826087\\
7.74448589900001	0.434782608695652\\
8.24868411700001	0.434782608695652\\
8.24868411700001	0.478260869565217\\
8.75395513599969	0.478260869565217\\
8.75395513599969	0.521739130434783\\
12.6146161039999	0.521739130434783\\
12.6146161039999	0.565217391304348\\
13.033184099	0.565217391304348\\
13.033184099	0.608695652173913\\
17.868712	0.608695652173913\\
17.868712	0.652173913043478\\
23.3582080000001	0.652173913043478\\
23.3582080000001	0.695652173913043\\
25.5570937519998	0.695652173913043\\
25.5570937519998	0.739130434782609\\
26.5043609089997	0.739130434782609\\
26.5043609089997	0.782608695652174\\
28.9125575679999	0.782608695652174\\
28.9125575679999	0.826086956521739\\
37.0544347870001	0.826086956521739\\
37.0544347870001	0.869565217391304\\
49.0503781360001	0.869565217391304\\
49.0503781360001	0.91304347826087\\
70.538559621	0.91304347826087\\
70.538559621	0.956521739130435\\
80.5708950009998	0.956521739130435\\
80.5708950009998	1\\
inf	1\\
};
\addlegendentry{Polynomial Regression};
\addplot [color=green,solid,line width=1.0pt]
table[row sep=crcr]{%
	-inf	0\\
	0	0\\
	0	0.0869565217391304\\
	10	0.0869565217391304\\
	10	0.217391304347826\\
	17.31	0.217391304347826\\
	17.31	0.260869565217391\\
	20	0.260869565217391\\
	20	0.304347826086957\\
	20.18	0.304347826086957\\
	20.18	0.347826086956522\\
	25.8	0.347826086956522\\
	25.8	0.391304347826087\\
	25.9	0.391304347826087\\
	25.9	0.434782608695652\\
	28.65	0.434782608695652\\
	28.65	0.478260869565217\\
	28.9	0.478260869565217\\
	28.9	0.521739130434783\\
	30	0.521739130434783\\
	30	0.608695652173913\\
	32.86	0.608695652173913\\
	32.86	0.652173913043478\\
	40	0.652173913043478\\
	40	0.695652173913043\\
	40.8	0.695652173913043\\
	40.8	0.739130434782609\\
	42.6	0.739130434782609\\
	42.6	0.782608695652174\\
	43.5	0.782608695652174\\
	43.5	0.826086956521739\\
	44.3	0.826086956521739\\
	44.3	0.869565217391304\\
	45.9	0.869565217391304\\
	45.9	0.91304347826087\\
	46.5	0.91304347826087\\
	46.5	0.956521739130435\\
	48.9	0.956521739130435\\
	48.9	1\\
	inf	1\\
};
\addlegendentry{Classification using SVM};

\end{axis}
\end{tikzpicture}%
	\vspace{-0.2cm}
	\caption{Localization error of regression vs fingerprinting}
	\vspace{-0.6cm}
	\label{pwrVspol}
\end{figure}

Fig. \ref{3rec} illustrates the average RSSI from transmitters placed at different distances. Obviously, the RSSI is decreasing with the distance. However, as shown in Fig. \ref{regr}, as we go further the influence of the distance on the RSSI starts decreasing. In addition to the RSSI measurements, Fig. \ref{regr} presents the output curve of polynomial regression and power series regression. An order three polynomial regression and a two-terms power series regression have been used. The output curve of the regression process has been used for distance estimation of the nodes in the testing phase. Fig. \ref{pwrVspol} shows the cumulative distribution function (CDF) of the location error for both polynomial and power series regression algorithms. It can be seen that polynomial regression provides slightly better performance than the power series regression. The figure also illustrates the difference between fingerprinting and a distance estimation approach. While distance estimation provides lower error with probability of 85\%, fingerprinting localization error is always limited by the distances between classes. Therefore, estimating the position of nodes in distances larger than 200m from the receiver leads to high errors ($>60m$). Thus, we limited the use of the distance estimation approach to classes with area of radius less than 200m.

\section{Conclusion}
This paper provides localization methods for UNB IoT networks by exploiting GPS nodes to estimate the location of other nodes. Two different implementations have been investigated. Firstly, only the communication over Sigfox has been considered. In this case we have proposed the use of fingerprinting of RSSI measurements where the meta-data of messages of the GPS nodes are used as a source for training data. This implementation provides the localization service for free, since the GPS nodes have to send their own messages anyway. On the other hand, enabling peer-to-peer communication can provide higher localization accuracy. This high accuracy can be achieved by using distance estimation for short distances. However, the cost comes as offline measurement necessity, with extra power and extra traffic only for the purpose of localization.
\section*{Acknowledgment}
The authors would like to thank the companies Quicksand and Engie-M2M for their cooperation.

%

\end{document}